\documentclass[aps,prl,twocolumn,showpacs,superscriptaddress,floatfix]{revtex4}

\usepackage{graphicx}
\usepackage{amssymb}

\begin{document}

\title{Ferromagnetic quantum critical point in URhGe doped with Ru}

\author{N. T. Huy}
\affiliation{Van der Waals - Zeeman Institute, University of
Amsterdam, Valckenierstraat~65, 1018 XE Amsterdam, The
Netherlands}
\author{A. Gasparini}
\affiliation{Van der Waals - Zeeman Institute, University of
Amsterdam, Valckenierstraat~65, 1018 XE Amsterdam, The
Netherlands}
\author{J. C. P. Klaasse}
\affiliation{Van der Waals - Zeeman Institute, University of
Amsterdam, Valckenierstraat~65, 1018 XE Amsterdam, The
Netherlands}
\author{A. de Visser}
\email{devisser@science.uva.nl} \affiliation{Van der Waals -
Zeeman Institute, University of Amsterdam, Valckenierstraat~65,
1018 XE Amsterdam, The Netherlands}
\author{S. Sakarya}
\affiliation{Department of Radiation, Radionuclides \& Reactors,
Delft University of Technology, Mekelweg 15, 2629 JB Delft, The
Netherlands}
\author{N. H. van Dijk}
\affiliation{Department of Radiation, Radionuclides \& Reactors,
Delft University of Technology, Mekelweg 15, 2629 JB Delft, The
Netherlands}

\date{\today}

\begin{abstract}
We have investigated the thermal, transport and magnetic
properties of URh$_{1-x}$Ru$_x$Ge alloys near the critical
concentration $x_{cr} = 0.38$ for the suppression of ferromagnetic
order. The Curie temperature vanishes linearly with $x$ and the
ordered moment $m_0$ is suppressed in a continuous way. At
$x_{cr}$ the specific heat varies as $c \sim TlnT$, the
$\gamma$-value $c/T|_{0.5K}$ is maximum and the temperature
exponent of the resistivity $\rho \sim T^n$ attains a minimum
value $n=1.2$. These observations provide evidence for a
ferromagnetic quantum phase transition. Interestingly, the
coefficient of thermal expansion and the Gr\"uneisen parameter
$\Gamma$ remain finite at $x_{cr}$ (down to $T = 1$~K), which is
at odds with recent scaling results for a metallic quantum
critical point.

\end{abstract}


\maketitle

In recent years interest has continued to grow in materials that
exhibit a quantum phase transition (QPT), $i.e.$ a transition at
zero temperature driven by quantum fluctuations
\cite{Sachdev-CUP-1999}. QPTs are fundamentally different from
their classical counter parts at finite $T$, where the transition
is due to thermal fluctuations of the order parameter. QPTs can be
induced in a wide range of materials, such as correlated metals
\cite{Schroeder-Nature-2000}, cuprate superconductors
\cite{vdMarel-Nature-2003}, common metals \cite{Yeh-Nature-2002}
and the two-dimensional electron gas \cite{Sondhi-RMP-1997}. This
is accomplished by adjusting a control parameter ($e.g.$ pressure
$p$, doping $x$, magnetic field $B$, or electron density) in order
to tune the system to a quantum critical point (QCP). At this
point the quantum critical fluctuations give rise to unusual
temperature laws (non-Fermi liquid behavior (nFL)) for the
magnetic, thermal and transport parameters \cite{Millis-PRB-1993},
and new collective states may emerge, $e.g.$ unconventional
superconducting \cite{Mathur-Nature-2001} or electronic states
\cite{Grigera-Science-2004}. This in turn calls for novel concepts
and theories
\cite{Schroeder-Nature-2000,Si-Nature-2001,Coleman-JPCM-2001}. In
order to provide a fruitful testing ground, it is important to
identify new systems and to investigate their critical behavior.

Strongly correlated electron systems, notably heavy-fermion
compounds based on the $f$-elements Ce, Yb or U, are especially
suited to study magnetic$-$to$-$non-magnetic QPTs, because the
ordering temperatures are low ($\sim10$~K) and the exchange
interaction can be modified relatively easily by an external
control parameter. Currently, there are two central questions that
are being addressed by studying QPTs in these materials. The first
issue is the fate of the quasiparticles when the antiferromagnetic
(AF) or ferromagnetic (FM) phase is entered. In the conventional
scenario a spin density wave is formed
\cite{Hertz-PRB-1976,Millis-PRB-1993} and the quasiparticles
preserve their itinerant character (as in CeIn$_{3-x}$Sn$_{x}$
\cite{Kuchler-PRL-2006}). Because the itinerant model is unable to
account for the nFL behavior in certain materials, an alternative
local quantum criticality model has been put forward
\cite{Schroeder-Nature-2000,Si-Nature-2001,Coleman-JPCM-2001}.
Here the quasiparticles (Kondo-screened moments) decompose at the
critical point in conduction electrons and local $f$-moments that
undergo magnetic order (as in CeCu$_{6-x}$Au$_{x}$
\cite{Schroeder-Nature-2000} and
YbRh${_2}$(Si$_{1-x}$Ge${_x}$)${_2}$ \cite{Custers-Nature-2003}).
The second captivating issue is the emergence of unconventional
superconducting (SC) states near the pressure induced QCPs in
CePd$_{2}$Si$_{2}$, CeIn$_{3}$ \cite{Mathur-Nature-2001} and
UGe${_2}$ \cite{Saxena-Nature-2000}. Evidence is at hand that in
these materials unconventional pairing is realized (d-wave pairing
for the AF and p-wave pairing for the FM systems). This strongly
suggests Cooper pairing mediated by AF or FM spin fluctuations
rather than by phonons. The coexistence of FM order and SC in
UGe$_{2}$ (and possibly in UIr \cite{Akazawa-JPCM-2004}) under
pressure, is uncommon in nature and attracts much attention.

In this paper we provide evidence for a ferromagnetic QPT in URhGe
doped with Ru. Our research is motivated by the unique properties
of the parent compound URhGe at ambient pressure: $(i)$ SC below
$T_{s} = 0.25$~K coexists with itinerant FM order (Curie
temperature $T_{C} = 9.5$~K) \cite{Aoki-Nature-2001}, and $(ii)$
re-entrant SC is induced by applying a large magnetic field ($B
\sim12$~T) \cite{Levy-Science-2005}. These observations
immediately prompted the question whether one can tune URhGe to a
FM QCP by mechanical or chemical pressure, with the objective to
probe the quantum critical fluctuations and possibly link these to
the SC pairing mechanism. Resistivity measurements under
hydrostatic pressure, however, revealed that $T_{C}$ {\em
increases} at a rate of $0.065$~K/kbar \cite{Hardy-PhysicaB-2005}.
Also, upon the application of uniaxial pressure $T_{C}$ increases
as was extracted from the Ehrenfest relation
\cite{Sakarya-PRB-2003}. As regards to chemical pressure, best
candidate dopants are Ru and Co, since among the neighboring
isostructural UTX compounds (T = transition metal and X = Ge or
Si) only URuGe and UCoGe have a paramagnetic ground state
\cite{Troc-JMMM-1988,Sechovsky-handbook-1998}. Indeed, FM order in
URhGe can be suppressed by replacing Rh by Ru and vanishes at 38
at.\% Ru \cite{Sakarya-PhysicaB-2006,Sakarya-CondMat-2006}. Here
we investigate the thermal, transport and magnetic properties of
URh$_{1-x}$Ru$_x$Ge alloys near the critical concentration $x_{cr}
= 0.38$. The observed nFL $T$ dependencies of the specific heat
and electrical resistivity, together with the smooth suppression
of the ordered moment, provide evidence for a continuous FM QPT.
This classifies URh$_{1-x}$Ru$_{x}$Ge as one of the scarce
$f$-electron systems in which a FM QCP can be reached by doping (a
FM QPT was also reported for CePd$_{1-x}$Rh$_{x}$
\cite{Sereni-PhysicaB-2005}, but here the transition is
"smeared").

Polycrystalline URh$_{1-x}$Ru$_x$Ge samples with $0.0 \leq x \leq
0.60$ were prepared  by arc-melting the constituents U, Rh, Ru
(all 3N) and Ge (5N) under a high-purity argon atmosphere in a
water-cooled copper crucible. The as-cast samples were wrapped in
Ta foil and annealed under high vacuum in quartz tubes for 10 days
at 875~$^\circ$C. Samples were cut by spark-erosion. Electron
probe micro analysis showed the single phase nature of the samples
within the resolution of 2\%. X-ray powder diffraction confirmed
the orthorhombic TiNiSi structure (space group $Pnma$)
\cite{Lloret-PhDthesis-1988,Prokes-PhysicaB-2002}. Upon
substituting Ru the unit cell volume $\Omega$ = 224.3~\AA$^3$ of
URhGe decreases linearly at a rate of 0.067~\AA$^3$ per at.\% Ru
(i.e. $\Delta\Omega$ = 1.1 \% at $x_{cr}$) in an anisotropic way,
the main effect being the reduction of the $a$ lattice parameter
\cite{Sakarya-CondMat-2006}.

The specific heat $c(T)$ was measured down to $0.4$ K using a
semi-adiabatic method in a home-built $^3$He system. Electrical
resistivity $\rho (T)$ data were collected in a commercial $^3$He
system (Heliox - Oxford Instruments, $T \geq 0.25$~K) using a low
frequency ac-resistance bridge. The thermal expansion $\alpha(T)$
was measured using a parallel-plate capacitance dilatometer in the
$T$ range 1-15 K. The dc magnetization $M(T)$ ($T \geq 1.8$~K) was
obtained using a Quantum Design SQUID magnetometer. Temperature
scans in magnetic fields $B$ up to 5 T were made after field
cooling.
\begin{figure}
\includegraphics[width=8cm]{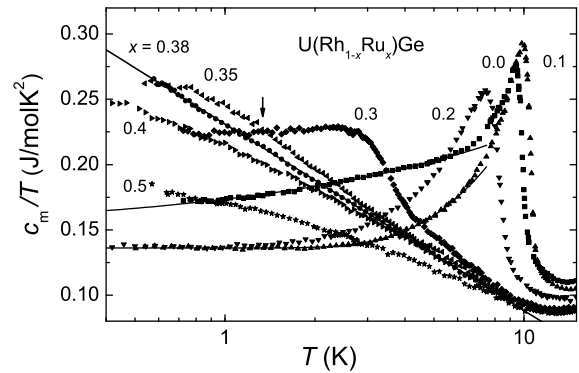} %Fig_URhRuGe_Specific_heat
\caption{$f$-electron specific heat of URh$_{1-x}$Ru$_x$Ge plotted
as $c_{m}/T ~vs~ \log T$ for $0 \leq x \leq 0.50$ as indicated.
For $x \leq 0.10$ the data are fitted to $c_{m}(T) = \gamma T +
\delta T^{3/2}e^{-\Delta /k_{B}T}$ with $\gamma$-, $\delta$- and
$\Delta /k_{B}$-values of $0.150,~ 0.146$ and $ 0.136$~J/molK$^2$,
$0.024,~ 0.041$ and $0.094$~J/molK$^{5/2}$ and $0, ~6.5$ and
$10.6$~K for $x = 0,~ 0.05$ and $0.10$, respectively (solid lines
for $x=0$ and $x=0.10$; data for $x=0.05$ not shown). The arrow
indicates $T_C$ for $x=0.35$. For $x_{cr} = 0.38$ $c_{m}/T \sim
\ln T$ over one and a half decade in $T$ (straight solid line).}
\end{figure}

The overall effect of Ru doping on ferromagnetism in URhGe is
presented in Fig.1, where we have plotted the $f$-electron
specific heat $c_{m}$, obtained after subtracting the lattice
contribution ($c_{lat}=\beta \, T^3$ for $T \leq 20$ K with $\beta
= 0.60 \cdot 10^{-3}$~J/mol\,K$^4$ \cite{Prokes-PhysicaB-2002}),
as $c_{m}/T ~vs~ \log T$ for $0 \leq x \leq 0.50$. Upon doping,
$T_{C}$ initially increases, but for $x \geq 0.10$ the ordering
peak shifts towards lower $T$ and weakens. Values of $T_{C}(x)$,
identified by the inflection points in $c/T ~vs~ T$ (on a linear
$T$ scale) at the high $T$ side of the peaks, are traced in Fig.2a
and are in excellent agreement with the values determined from
$M(T)$ and $\rho(T)$ \cite{Sakarya-PhysicaB-2006}. For $x \geq
0.20$ $T_{C}$ decreases linearly with $x$ at a rate of
$0.45$~K/at.\% Ru. For $x=0$ the magnetic specific heat for $T
\leq 5$~K is described by $c_{m}(T) = \gamma T + \delta T^{3/2}$,
where $\gamma$ is the linear coefficient of the electronic
specific heat and the second term is the spin wave contribution
\cite{Tari-ICP-2003}. The values for $\gamma$ and $\delta$
extracted by fitting the data (see Fig.1) are in good agreement
with the values reported in Ref.\cite{Prokes-PhysicaB-2002}. Upon
doping Ru an energy gap $\Delta$ opens in the magnon spectrum and
the specific heat for $x=0.05$ and $0.10$ now follows the relation
($T \leq 5$~K) $c_{m}(T) = \gamma T + \delta T^{3/2}e^{-\Delta
/k_{B}T}$ \cite{Tari-ICP-2003} (see fits in Fig.1). The most
important result of our specific heat experiments however is the
pronounced $c_{m}(T) = -bT\ln ( T/T_{0} )$ dependence for
$x_{cr}$, where $b = 0.062$~J/mol\,K$^2$ and $T_0$ = 41~K. This
nFL term is observed over one and a half decade in $T$
($0.5-9$~K). At $x_{cr}$ $c/T|_{0.5K}(x)$ has a maximum (Fig.2c).
The total $f$-electron entropy obtained by integrating $c_{m}/T
~vs~ T$ between 0.5 and $\sim 15$~K amounts to $\sim 0.48R\ln 2$
for $x=0$ and decreases to $0.33R\ln 2$ at $x_{cr}$. Its small
value confirms the itinerant nature of the FM transition (the
ordered moment $m_{0}$ is $0.4 ~\mu _{B}$ for $x=0$
\cite{Prokes-PhysicaB-2002,Aoki-Nature-2001}).
\begin{figure}
\includegraphics[width=8cm]{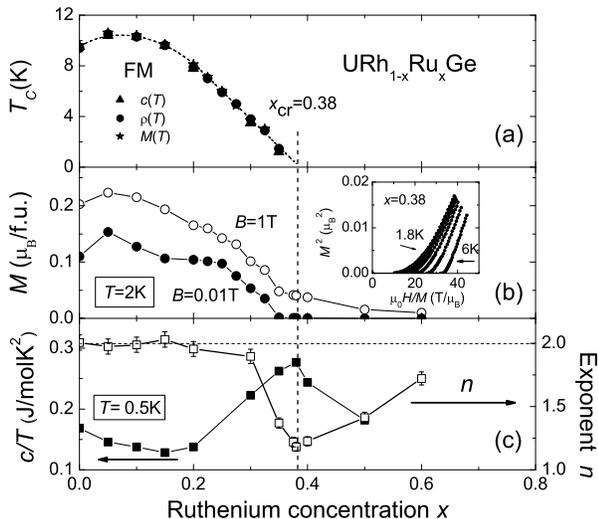} %Fig_URhRuGe_Concentration_dependence
\caption{(a) Curie temperature of URh$_{1-x}$Ru$_{x}$Ge determined
from $c(T)$ ($\blacktriangle$), $\rho (T)$ ($\bullet$) and $M(T)$
($\bigstar$). The critical Ru content is $x_{cr}=0.38$ (vertical
dashed line). (b) Magnetization $M$ at 2 K in $B=0.01$ ($\bullet$)
and 1 T ($\circ$). Inset: Arrott plot for $x=0.38$ at 1.8 K $\leq
T \leq 6$~K . (c) $c/T$ at $T=0.5$~K ($\blacksquare$) and the
exponent $n$ ($\square$) of $\rho \sim T^{n}$. The horizontal
dashed line indicates $n=2$.}
\end{figure}

The electrical resistivity of URh$_{1-x}$Ru$_x$Ge ($x \leq 0.60$)
at high $T$ \cite{Sakarya-CondMat-2006} shows the behavior typical
for a FM Kondo-lattice. The data for $x=0.38$ are shown in the
inset in Fig.3, where the maximum near 130 K signals the formation
of the Kondo-lattice. For the FM compounds at low $T$ a kink in
$\rho(T)$ (and maximum in $d\rho(T) /dT$) marks $T_{C}$. For all
doped samples the total resistivity drop in the $T$ interval 0-300
K is $\sim$150-250$~\mu \Omega$cm, which is usual for uranium
intermetallics \cite{Sechovsky-handbook-1998}. However, the
residual resistivity values $\rho_{0}$ are large ($\sim$
200-300$~\mu \Omega$cm), which is due to the brittleness of the
samples (cracks). Consequently, the RRR values ($R
(300$K)/R($0$K)) are small ($\sim 2$). In Fig.3 we show $\rho(T)$
at low $T$ for $0.10 \leq x \leq 0.60$. For a FM with gapped
magnon modes $\rho(T) = \rho_{0}+ AT^{n} + BT \Delta e^{-\Delta
/k_{B}T}(1+2k_{B}T/\Delta)$ \cite{Andersen-PRB-1979}, where the
2nd term is the electron-electron scattering term ($i.e.$ the FL
term when $n=2$) and the 3rd term yields the scattering from
magnons. For $x=0.10$ and $0.20$ fits reveal that the 2nd term is
dominant ($A \gg B$) and $\rho(T) \sim T^{2.0 \pm 0.1}$ over a
wide $T$ range in the FM state (see Fig.3). Therefore, we conclude
that scattering from magnons can be neglected in our
polycrystalline samples and we restrict the analysis to fitting
$\rho(T)= \rho_{0}+AT^{n}$ (see Fig.3). The values of $n$
extracted (by taking the best fit over the largest $T$ interval)
are shown in Fig.2c. $n(x)$ attains a minimum value $n=1.2$ at
$x_{cr}$, followed by a slow recovery to the FL value $n=2$ there
above.
\begin{figure}
\includegraphics[width=7cm]{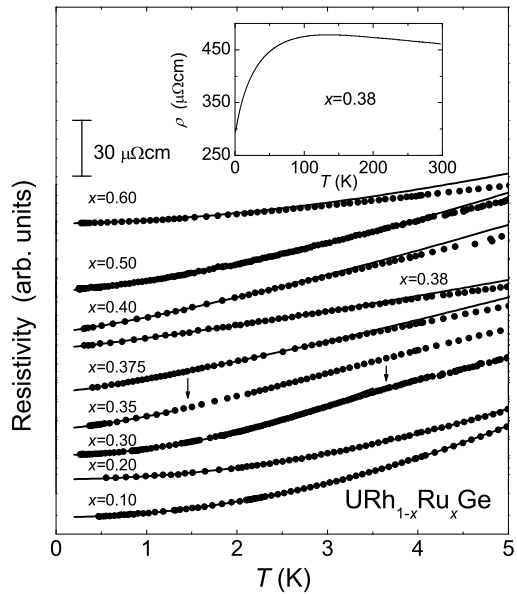} %Fig_URhRuGe_Resistivity
\caption{Resistivity of URh$_{1-x}$Ru$_{x}$Ge for $0.10 \leq x
\leq 0.60$. The bar gives the absolute scale. The arrows for
$x=0.30$ and $0.35$ indicate $T_{C}$ obtained from additional data
sets. The solid lines are fits to $\rho (T) = \rho_{0} + A T^{n}$.
For $x \leq 0.3$ $n=2.0 \pm 0.1$. For $x_{cr}=0.38$ $n=1.2$ is
minimum. Inset: Resistivity for $x=0.38$ up to 300 K.}
\end{figure}

The magnetization $M(T)$ for all samples was measured in
$B=0.01$~T and $1$~T down to $1.8$~K. In addition $M(B)$ was
measured at fixed $T$ in order to produce Arrott plots ($M^{2}~
{\emph vs}~ B/M$). $M|_{2K}$-values are traced in Fig.2b. For pure
URhGe $M|_{2K}$ in $1$~T $\simeq~ 0.2~\mu_{B}$ in agreement with
the polycrystalline average $\frac{1}{2}m_{0}$ for a uniaxial FM
($m_{0}=0.4~\mu_{B}$ directed along the $c$-axis
\cite{Aoki-Nature-2001}). In $0.01$~T a reduced value $M|_{2K}
\simeq 0.11~\mu_{B}$ is observed due to demagnetizing effects.
Values of $T_{C}$ (Fig.2a) were determined from the inflection
points in $M(T)$ in $0.01$~T and from the Arrott plots. For $x
\geq 0.38$ the Arrott plots ($T \geq 1.8$~K) no longer indicate
magnetic order (see inset in Fig.2b for $x=0.38$). A most
important feature of the data is the gradual decrease of
$M|_{2K}$$(x)$. For $B=0.01$~T $M|_{2K}(x)$ smoothly goes to $0$
at $x=0.35~ (T_{C}= 1.3 \pm 0.1$~K), while for $B=1$~T a finite
field induced $M|_{2K}$ remains. We conclude that the
FM-paramagnetic transition as a function of $x$ is a continuous
(2nd order) phase transition.
\begin{figure}
\includegraphics[width=8cm]{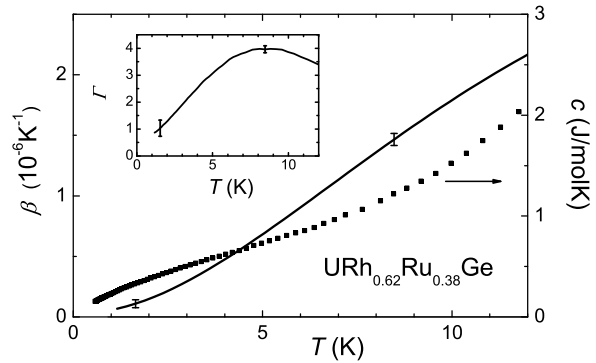} %Fig_URhRuGe_Alpha_Gamma
\caption{Coefficient of volume thermal expansion $\beta (T)$
(solid line) and specific heat $c(T)$ ($\bullet$) of
URh$_{0.62}$Ru$_{0.38}$Ge. Inset: Gr\"{u}neisen ratio $\Gamma$ as
a function of $T$.}
\end{figure}

In Fig.4 we show the coefficient of volume thermal expansion
$\beta(T)$ for $x_{cr}=0.38$ at $T \geq 1$~K. The data (solid
line) is obtained by averaging $\alpha_{i}(T)$ measured for three
orthogonal directions on the polycrystalline sample ($\beta=
\Sigma_{i} \alpha_{i}$) in order to eliminate possible anisotropy
effects due to crystallites with preferred orientations. The $T$
dependence of $\beta$ at low $T$ is weaker than that of the
specific heat (see Fig.4). Concurrently, the Gr\"{u}neisen ratio
$\Gamma = V_{m}\beta / \kappa c$ decreases below $T \sim 7$~K
(here the molar volume $V_{m} = 3.36 \times 10^{-5}$ m$^3$/mol and
isothermal compressibility $\kappa \simeq 10^{-11}$~Pa$^{-1}$
\cite{Sakarya-CondMat-2006}). The quasi-linear behavior of
$\Gamma(T)$ for 1 K $\leq T \leq 5$ K suggests an unusual $T$
variation of $\beta$, $i.e.$ roughly proportional to $T^{2}\ln T$.

Having documented the critical behavior of the
URh$_{1-x}$Ru$_{x}$Ge alloys we conclude that our $c(T)$, $\rho
(T)$ and $M(T)$ data provide evidence for a continuous FM QPT with
$x_{cr} = 0.38$. The most compelling evidence is the specific heat
$c_{cr} \sim T \ln (T/T_{0})$ observed over one and a half decade
in $T$ (Fig.1) \cite{Millis-PRB-1993} and the concomitant maximum
in $c/T|_{0.5K}(x)$ (Fig.2c). The temperature $T_{0}= 41$~K is
large, which indicates that our $c(T)$ experiments down to
$T=0.4$~K ($T/T_{0} \simeq 0.01$) indeed probe the quantum
critical regime. It will be interesting to investigate whether the
$c/T \sim lnT$ behavior persists even at lower $T$. Eventually,
however, $c/T$ will saturate because of crystallographic disorder
inherent to the URh$_{1-x}$Ru$_{x}$Ge alloys. Further support for
a QCP is provided by the critical behavior in the resistivity
$\rho_{cr} \sim T^{1.2}$ up to 2 K. The exponent $n(x)$ has a
pronounced minimum at $x_{cr}$ (Fig.2c). The value $n=1.2$ is
smaller than the value $n=5/3$ predicted for a clean FM QCP
\cite{Moriya-Book-1985}. This is not unexpected as disorder
reduces $n$ \cite{Pfleiderer-Nature-2001}. The itinerant nature of
the FM state and the smooth suppression of $m_{0}$ pointing to a
continuous phase transition, strongly suggest that the QPT in
URh$_{1-x}$Ru$_x$Ge is of the Hertz-Millis type
\cite{Hertz-PRB-1976,Millis-PRB-1993}, albeit with modified
exponents due to the effects of doping (notably emptying the
$d$-band and alloy disorder). For instance, for an itinerant clean
FM QPT one expects $T_{C} \sim (x_{c} - x)^{3/4}$ (dimension
$d=3$, dynamical critical exponent $z=3$), while we obtain $T_{C}
\sim (x_{c} - x)$ over a wide range $0.20 \leq x \leq 0.35$.
Deviations from the clean behavior are also observed in
$f$-electron materials with a pressure induced continuous FM QPT,
like CeSi$_{1.81}$ \cite{Drotziger-PRB-2006}. On the other hand,
for $d$-electron alloys with a continuous FM QPT ($e.g.$
Ni$_{x}$Pd$_{1-x}$ \cite{Nicklas-PRL-1999} and
Zr$_{1-x}$Nb$_{x}$Zn$_{2}$ \cite{Sokolov-PRL-2006}) the data are
to a large extent in agreement with the itinerant model. Further
theoretical work is required to clarify these issues.

Finally, we discuss our results for the thermal expansion and the
Gr\"{u}neisen parameter. The finite $\Gamma$-value at low $T$ is
at variance with the recent prediction of a diverging
Gr\"{u}neisen ratio $\Gamma \sim T^{-1/z\nu}$ at the QCP ($\nu$ is
the correlation length exponent) \cite{Zhu-PRL-2003}. For the case
of an itinerant FM QCP the scaling results are $\beta_{cr} \sim
T^{1/3}$ and $c_{cr} \sim T \log (1/T)$, whence $\Gamma_{cr} \sim
\beta_{cr}/c_{cr} \sim ((T^{2/3} \log(1/T))^{-1}$
\cite{Zhu-PRL-2003}. While the specific heat follows the expected
behavior, the thermal expansion clearly does not ($\beta \sim
T^{2}\ln T$ for 1 K $\leq T \leq 5$~K). With the value
$T_{0}=41$~K extracted from $c_{cr}$ we calculate that
$\Gamma_{cr}$ within the scenario of Ref.\cite{Zhu-PRL-2003}
should have a minimum near $8$~K and diverge at lower $T$. This is
obviously not the case experimentally (Fig.4). The only other
system for which the Gr\"{u}neisen ratio near a FM QPT has been
investigated so far is CePd$_{1-x}$Rh${_x}$
\cite{Sereni-PRB-2007}. In this system a non-diverging
($T$-independent) $\Gamma$ was also observed in the critical
regime.

In conclusion, we have investigated the thermal, transport and
magnetic properties of URh$_{1-x}$Ru$_x$Ge near the critical
concentration for the suppression of FM order. At $x_{cr}=0.38$ $c
\sim T \ln T$, the $\gamma$-value $c/T|_{0.5K}$ has a maximum and
the $T$ exponent in the resistivity attains the nFL value $n=1.2$.
Together with the gradual suppression of the ordered moment
$m_{0}$ the data provide evidence for a continuous FM quantum
phase transition. This offers the sole opportunity thus far to
investigate FM spin fluctuations in URhGe under quantum critical
conditions. The identification of the FM QCP at ambient pressure
in URhGe doped with Ru paves the road to a host of experiments on
this unique material.

This work was part of the research program of FOM (Dutch
Foundation for Fundamental Research of Matter) and COST Action P16
ECOM.


\begin{thebibliography}{99}


\bibitem{Sachdev-CUP-1999} See $e.g.$ S. Sachdev,
{\it Quantum Phase Transitions} (Cambridge University Press,
Cambridge, England, 1999).

\bibitem{Schroeder-Nature-2000} A. Schr\"{o}der, G. Aeppli, R. Coldea, M. Adams,
O. Stockert, H. v. L\"{o}hneysen, E. Bucher, R. Ramazashvili, and
P. Coleman, Nature (London) {\bf 407}, 351 (2000).


\bibitem{vdMarel-Nature-2003} D. van der Marel, H. J. A. Molegraaf, J. Zaanen,
Z. Nussinov, F. Carbone, A. Damascelli, H. Eisaki, M. Greven, P.
H. Kes, and M. Li, Nature (London) {\bf 425}, 271 (2003).


\bibitem{Yeh-Nature-2002} A. Yeh, Yeong-Ah Soh, J. Brooke, G. Aeppli, T. F. Rosenbaum, and S. M. Hayden,
Nature (London) {\bf 419}, 459 (2002).


\bibitem{Sondhi-RMP-1997} S. L. Sondhi, S. M. Girvin, J. P. Carini, and D. Shahar,
Rev. Mod. Physics {\bf 69}, 315 (1997).


\bibitem{Hertz-PRB-1976} J. Hertz,
Phys. Rev. B {\bf 14}, 1165 (1976).

\bibitem{Millis-PRB-1993} A. J. Millis,
Phys. Rev. B {\bf 48}, 7183 (1993).

\bibitem{Mathur-Nature-2001} N. D. Mathur, F. M. Grosche, S. R. Julian, I. R. Walker, D. M. Freye, R. K. W.
Haselwimmer, and G. G. Lonzarich, Nature (London) {\bf 394}, 39
(1998).


\bibitem{Grigera-Science-2004} S. A. Grigera, P. Gegenwart, R. A. Borzi, F. Weickert,
A. J. Schofield, R. S. Perry, T. Tayama, T. Sakakibara, Y. Maeno,
A. G. Green, and A. P. Mackenzie, Science {\bf 306}, 1154 (2004).


\bibitem{Si-Nature-2001} Q. Si, S. Rabello, K. Ingersent, and J. L. Smith,
Nature (London) {\bf 413}, 804 (2001).


\bibitem{Coleman-JPCM-2001} P. Coleman, C. P{\'e}pin, Q. Si, and R. Ramazashvili,
J. Phys.: Condens. Matter {\bf 13}, R723 (2001).


\bibitem{Kuchler-PRL-2006} R. K{\"u}chler, P. Gegenwart, J. Custers, O. Stockert, N. Caroca-Canales,
C. Geibel, J.G. Sereni, and F. Steglich, Phys. Rev. Lett. {\bf
96}, 256403 (2006).


\bibitem{Custers-Nature-2003} J. Custers, P. Gegenwart, H. Wilhelm, K. Neumaier, Y. Tokiwa,
O. Trovarelli, C. Geibel, F. Steglich, C. P{\'e}pin, and P.
Coleman, Nature (London) {\bf 424}, 524 (2003).


\bibitem{Saxena-Nature-2000} S. S. Saxena, P. Agarwal, K. Ahilan, F. M. Grosche, R. K. W. Haselwimmer, M. J. Steiner,
E. Pugh, I. R. Walker, S. R. Julian, P. Monthoux, G. G. Lonzarich,
A. Huxley, I. Sheikin, D. Braithwaite, and J. Flouquet, Nature
(London) {\bf 406}, 587 (2000).


\bibitem{Akazawa-JPCM-2004} T. Akazawa, H. Hidaka, T. Fujiwara, T. C. Kobayashi, E. Yamamoto,
Y. Haga, R. Settai, and Y. {\={O}}nuki, J. Phys.: Condens. Matter
{\bf 16}, L29 (2004).

\bibitem{Aoki-Nature-2001} D. Aoki, A. Huxley, E. Ressouche, D. Braithwaite, J. Flouquet,
J. P. Brison, E. Lhotel, and C. Paulsen, Nature (London) {\bf
413}, 613 (2001).


\bibitem{Levy-Science-2005} F. L\'{e}vy, I. Sheikun, B. Grenier, and A. D. Huxley,
Science {\bf 309}, 1343 (2005).
%

\bibitem{Hardy-PhysicaB-2005} F. Hardy, A. Huxley, J. Flouquet, B. Salce, G. Knebel, D. Braithwaite,
D. Aoki, M. Uhlarz, and C. Pfleiderer, Physica B {\bf 359-361},
1111 (2005).


\bibitem{Sakarya-PRB-2003} S. Sakarya, N. H. van Dijk, A. de Visser, and E. Br{\"{u}}ck,
Phys. Rev. B {\bf 67}, 144407 (2003).


\bibitem{Troc-JMMM-1988} R. Tro{\'{c}} and V. H. Tran,
J. Magn. Magn. Mater. {\bf 73}, 389 (1988).

\bibitem{Sechovsky-handbook-1998} V. Sechovsk{\'{y}} and L.
Havela, {\it Handbook of Magnetic Materials} Vol.~11 ed. K. H. J.
Buschow (North Holland, Amsterdam, 1998) pp.~1-289.

\bibitem{Sakarya-PhysicaB-2006} S. Sakarya, N. H. van Dijk, N. T. Huy, and A. de Visser,
Physica B {\bf 378-380}, 970 (2006).


\bibitem{Sakarya-CondMat-2006} S. Sakarya, N. T. Huy, N. H. van Dijk, A. de Visser, M. Wagemaker,
A. C. Moleman,  T. J. Gortenmulder, J. C. P. Klaasse, M. Uhlarz,
and H. v. L{\"o}hneysen, J. Alloys and Compounds, in print;
e-print cond-mat/0609557.


\bibitem{Sereni-PhysicaB-2005} J.G. Sereni, R. K{\"u}chler, and C.
Geibel, Physica B {\bf 359-361}, 41 (2005).

\bibitem{Prokes-PhysicaB-2002} K. Proke\u{s}, T. Tahara, Y. Echizen, T. Takabatake, T. Fujita, I. H. Hagmusa,
J. C. P. Klaasse, E. Br\"{u}ck, F. R. de Boer, M. Divi\u{s}, and
V. Sechovsk{\'{y}}, Physica B {\bf 311}, 220 (2002).


\bibitem{Lloret-PhDthesis-1988} B. Lloret, {\it Ph.D. Thesis},
(University of Bordeaux I, 1988).

\bibitem{Tari-ICP-2003} See $e.g.$ A. Tari, {\it The Specific Heat of
Matter at Low Temperatures} (Imperial College Press, London,
2003).

\bibitem{Andersen-PRB-1979} N. Hessel Andersen, and H. Smith,
Phys. Rev. B {\bf 19}, 384 (1979).

\bibitem{Moriya-Book-1985} T. Moriya, {\it Spin Fluctuations in
Itinerant Electron Magnets} (Springer Verlag, Berlin, 1985).

\bibitem{Pfleiderer-Nature-2001} See e.g. C. Pfleiderer, S. R. Julian, and G. G.
Lonzarich, Nature (London) {\bf 414}, 427 (2001).

\bibitem{Drotziger-PRB-2006} S. Drotziger, C. Pfleiderer, M. Uhlarz, H. v. L{\"o}hneysen,
D. Souptel, W. L{\"o}ser, and G. Behr, Phys. Rev. B {\bf 73},
214413 (2006).


\bibitem{Nicklas-PRL-1999} M. Nicklas, M. Brando, G. Knebel, F. Mayr, W. Trinkl, and A. Loidl,
Phys. Rev. Lett. {\bf 82}, 4268 (1999).


\bibitem{Sokolov-PRL-2006} D. A. Sokolov, M. C. Aronson, W. Gannon, and Z. Fisk,
Phys. Rev. Lett. {\bf 96}, 116404 (2006).


\bibitem{Zhu-PRL-2003} L. Zhu, M. Garst, A. Rosch, and Q. Si,
Phys. Rev. Lett. {\bf 91}, 66404 (2003).


\bibitem{Sereni-PRB-2007} J. G. Sereni, T. Westerkamp, R. K\"{u}chler, N. Caroca-Canales, P. Gegenwart, and C. Geibel,
Phys. Rev. B {\bf 75}, 024432 (2007).



\end{thebibliography}
\end{document}